\documentclass[review]{elsarticle}

\makeatletter
\def\ps@pprintTitle{%
 \let\@oddhead\@empty
 \let\@evenhead\@empty
 \def\@oddfoot{\centerline{\thepage}}%
 \let\@evenfoot\@oddfoot}
\makeatother

\usepackage[margin=1.2in,footskip=0.6in]{geometry}
\usepackage{lineno,hyperref,amsmath,amssymb}
\usepackage{subfig}
\usepackage[table]{xcolor}
\usepackage{url}
\definecolor{Gray}{gray}{0.9}

\begin{document}

\begin{frontmatter}

\title{Gain suppression mechanism observed in Low Gain Avalanche Detectors}

%% Group authors per affiliation:
\cortext[mycorrespondingauthor]{Corresponding author}
\author[a]{E. Curr\'{a}s\corref{mycorrespondingauthor}}
\ead{ecurrasr@cern.ch}
\author[a,b]{M. Fern\'{a}ndez}
\author[a]{M. Moll}

\address[a]{CERN, Organisation europ\'{e}nne pour la recherche nucl\'{e}aire, CH-1211 Geneva 23, Switzerland}
\address[b]{Instituto de F\'{i}sica de Cantabria (CSIC-UC), Avda. los Castros s/n, E-39005 Santander, Spain}
%% or include affiliations in footnotes:
\begin{abstract}

Low Gain Avalanche Detectors (LGADs) is one of the candidate sensing technologies for future 4D-tracking applications and recently have been qualified to be used in the ATLAS and CMS timing detectors for the CERN High Luminosity Large Hadron Collider upgrade. LGADs can achieve an excellent timing performance by the presence of an internal gain that improves the signal-to-noise ratio leading to a better time resolution.

These detectors are designed to exhibit a moderate gain with an increase of the reverse bias voltage. The value of the gain strongly depends on the temperature. Thus, these two values must be kept under control in the experiments to maintain the gain within the required values. A reduction in the reverse bias or an increase in the temperature will reduce the gain significantly.

In this paper, a mechanism for gain suppression in LGADs is going to be presented. It was observed, that the gain measured in these devices depends on the charge density projected into the gain layer, generated by a laser or a charged particle in their bulk. Measurements performed with different detectors showed that ionizing processes that induce more charge density in the detector bulk lead to a decrease in the detector’s measured gain.

Measurements conducted with IR-laser and Sr-90 in the lab, modifying the charge density generated in the detector bulk, confirmed this mechanism and will be presented here.

\end{abstract}

\begin{keyword}
\texttt LGAD, gain, charge collection, timing detector, jitter, charge density, screening effect.
\end{keyword}

\end{frontmatter}

%\linenumbers

\section{Introduction}

\begin{figure}[!t]
\centering
\includegraphics[width=0.90\textwidth]{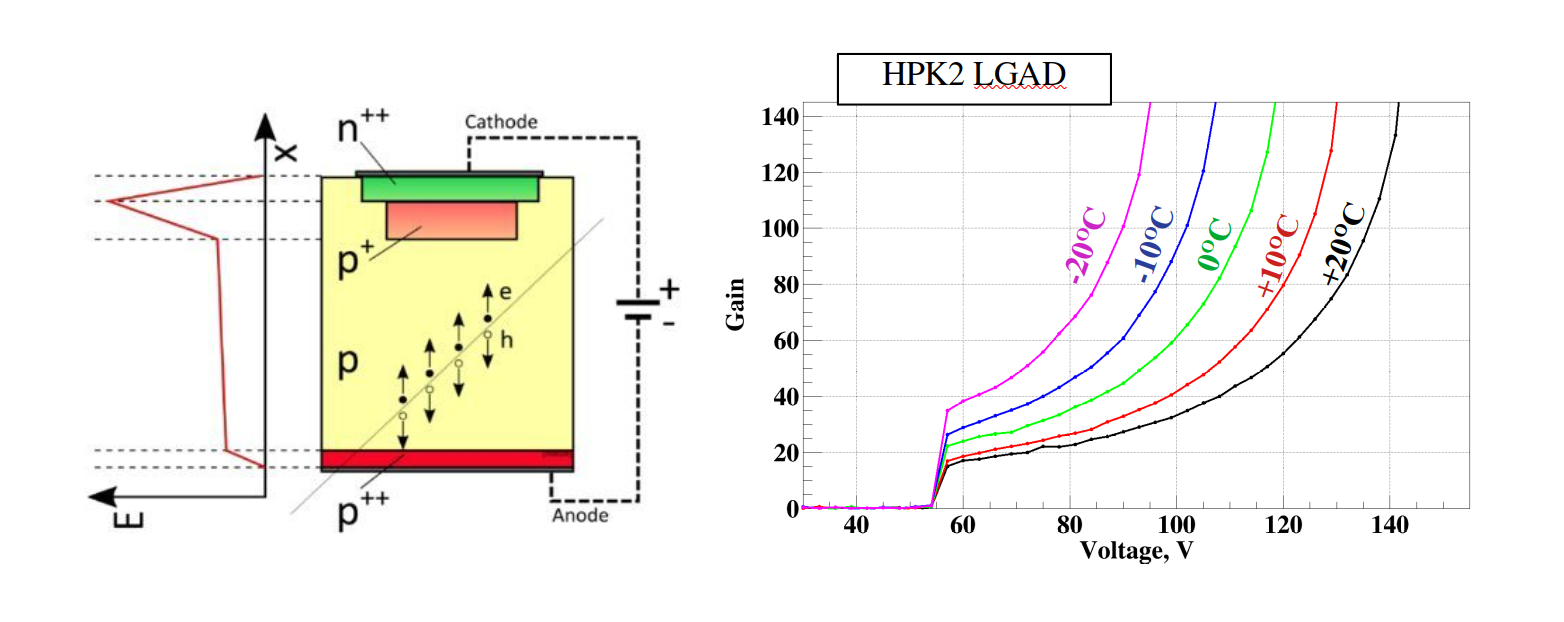}
\caption{On the left: a sketch of the cross-section of a pad-like LGAD with a charged particle passing through. A qualitative profile of the electric field strength is shown too, where the peak is located in the gain layer region ($p^+$) in which the avalanche happens. On the right: gain as a function of the voltage for a $50\,\mu m$  LGAD measured at five different temperatures.} 
\label{LGAD_intro}
\centering
\end{figure}

The Low Gain Avalanche Detector (LGAD) technology was pioneered within the RD50 collaboration \cite{PELLEGRINI201412} and consists of an n-on-p silicon detector with internal gain. The internal gain is obtained by adding a highly doped p-layer below the p-n junction of the detector's PIN diode structure. This highly doped region creates locally a very high electric field strength when the detector is set under bias. This electric field induces an avalanche multiplication of the electrons passing the high field region and thus creates additional electron-hole pairs. The structure is designed to exhibit a moderate gain with a smooth increase of the gain over a wide range of reverse bias voltage values. A schematic sketch of the cross-section of a standard pad-like LGAD is shown on the left side of figure\,\ref{LGAD_intro}. The LGAD technology has been qualified for the use in the MIP Timing Detector of the CMS experiment and in the High-Granularity Timing Detector (HGTD) of the ATLAS experiment for HL-LHC operations \cite{CERN-LHCC-2017-027,CERN-LHCC-2018-023}.

To maximize the performance of LGAD detectors, the gain at which they have to operate must be carefully tuned to achieve the desired time resolution. If the gain varies during operation in the experiments, the time resolution is going to be affected which can degrade and compromise the performance of the whole experiment. For this reason, the reverse bias voltage and the temperature at which the detectors are operated must be always kept under control. Also, during the HL-LHC operation or any other operation in strong radiation fields, the radiation damage will progressively damage the detectors, and in consequence, the operation voltage must be properly adjusted with increasing radiation damage \cite{Curras:Vertex}. In figure \ref{LGAD_intro} right, the gain dependence on the reverse bias, measured at five different temperatures for a $50\,\mu m$ thick HPK LGAD, is shown. The variation of gain, and in consequence also the shift of the breakdown voltage to lower values, is clearly visible demonstrating the need to adapt the operational voltage to environmental conditions.

\bigskip

In this paper, a further mechanism that affects the gain during the operation of the detectors is going to be presented. It will be shown that the gain depends on the ionization density of the process leading to the charge deposition in the sensor bulk. For this, signals obtained by exposure to laser pulses of different intensity and beta particles are compared in the following sections. To the best of our knowledge, these differences in gain were not reported yet and although there are many studies about similar effects, like the plasma effect in silicon since it was first noted \cite{Miller:4315762}, there are no studies about the influence of the ionizing charge density generation below the plasma regime in the impact ionizing process in silicon. This 'gain damping effect' should be taken into account for the operation of LGADs in experiments because depending on the measurement conditions the gain can change significantly. The paper concludes with a qualitative explanation of the mechanism leading to the effect.

\section{Set-ups and samples}

The samples under study in this paper are LGADs and PINs produced by Hamamatsu Photonics (HPK) and Centro Nacional de Microelectr\'{o}nica (CNM-IMB). The LGAD and PIN sensors from the respective producers differ only in the addition of the p$^+$-implant ({\em gain layer}) for LGADs.  The HPK samples were from the production run S10938-6130 (also called HPK prototype 2 or HPK2) produced on a wafer with a $50\,\mu m$ epitaxial layer on a $150\,\mu m$ thick low resistivity support wafer. The CNM samples were from the production run 12916, they were produced on a $50\,\mu m$ Float Zone wafer bonded to a $300\,\mu m$ low resistivity Czochralski wafer as support. 

\bigskip

\begin{figure}[!t]
\centering
\includegraphics[width=0.95\textwidth]{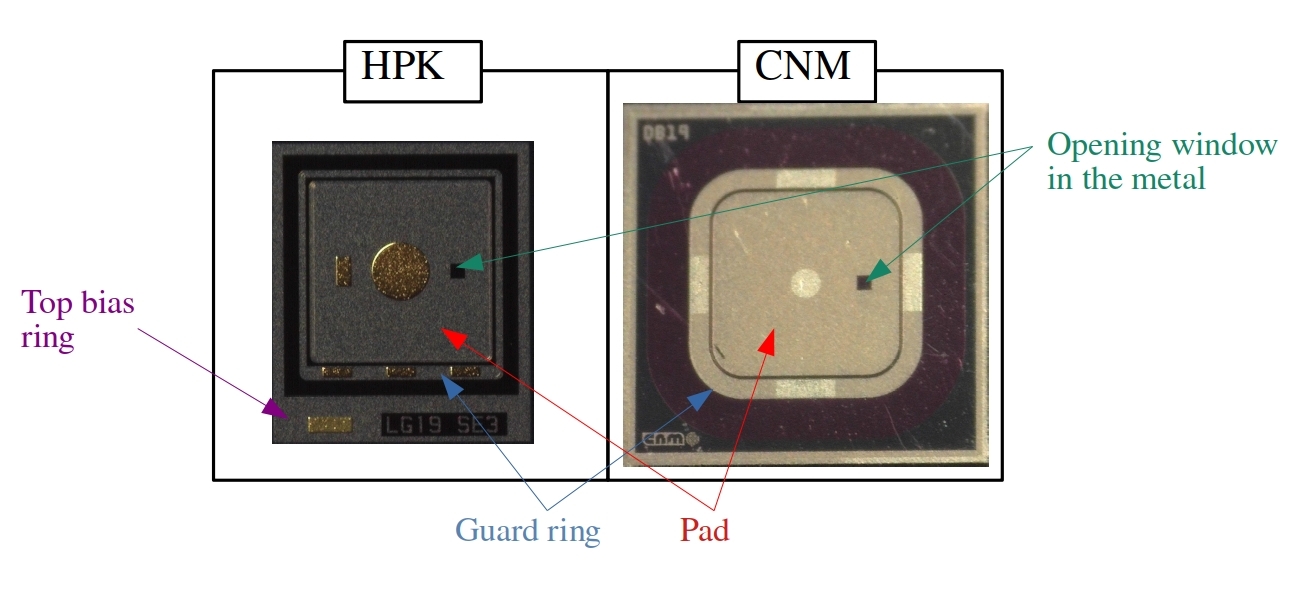}
\caption{Photographs (not scaled) of the two LGAD types characterized in this work. The left picture shows a top view of a device produced by HPK and the right one the top view of an LGAD produced by CNM. The corresponding PIN sensors of the two producers look identical.} 
\label{LGAD_pics}
\centering
\end{figure}

\begin{figure}[!t]
\centering
\includegraphics[width=0.45\columnwidth]{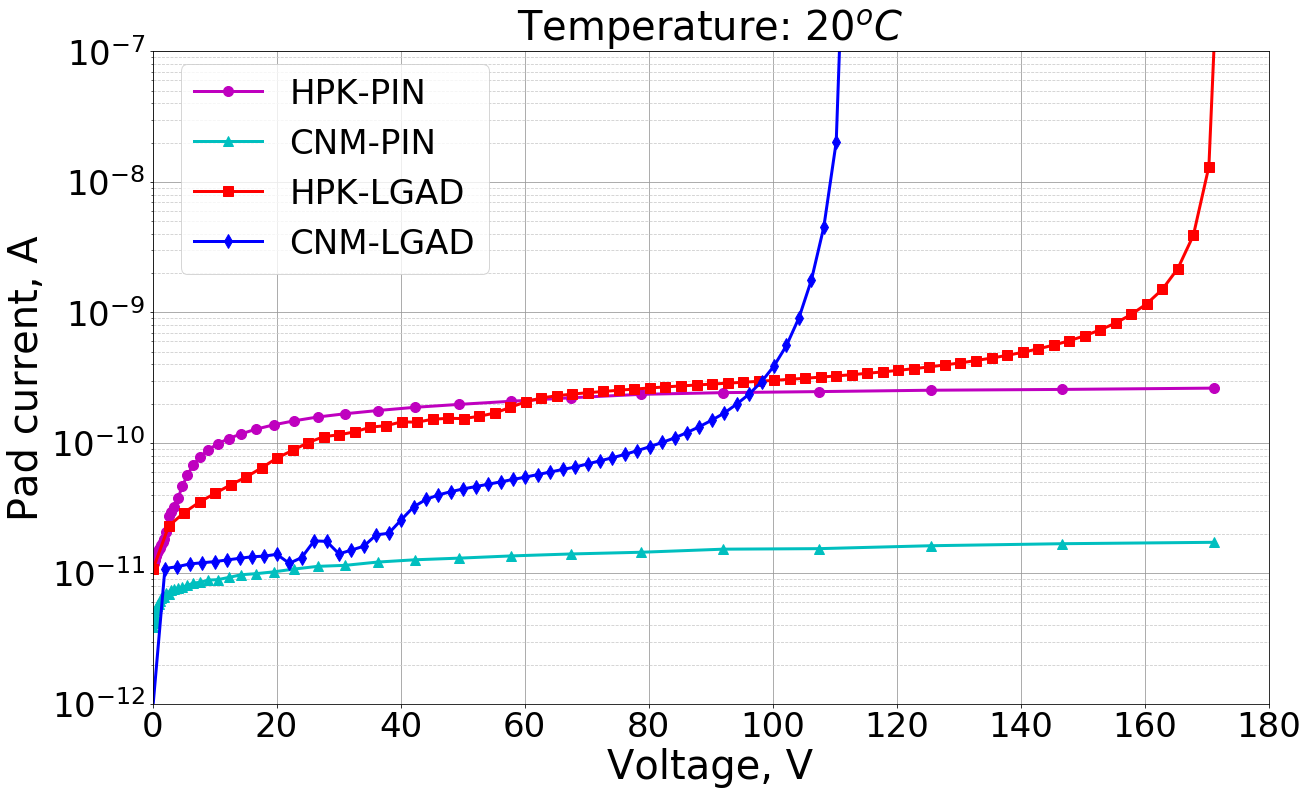}\hfil
\includegraphics[width=0.45\columnwidth]{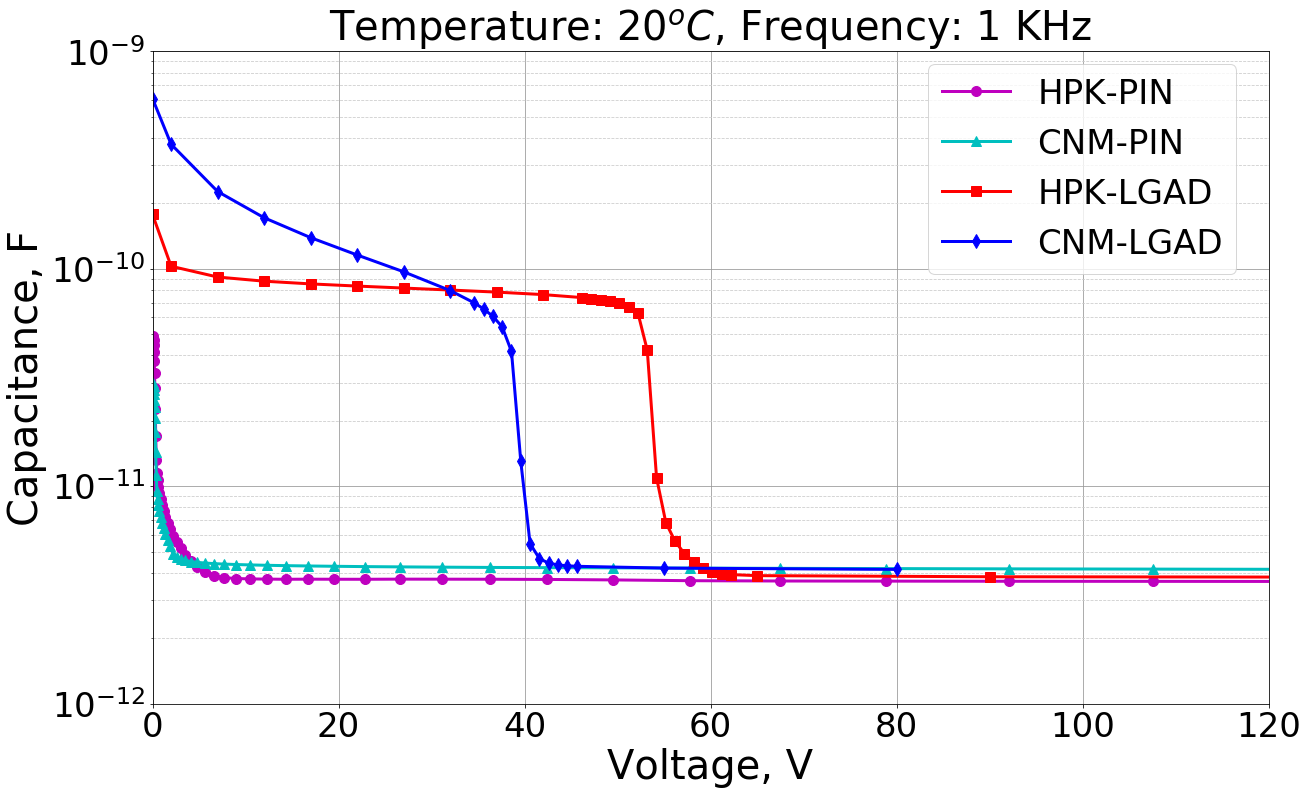}\hfil
\caption{Electrical characterization of the samples under-test in this paper. The left plot shows the pad current as a function of the reverse bias, and the right one shows the capacitance as a function of the reverse bias.} 
\label{LGADs_IVCV}
\centering
\end{figure}

All samples have an active area of $1.3\times1.3\,mm^2$ and a guard ring structure surrounding the pad. To allow laser illumination from the pad side (i.e. front electrode), they have an opening window  of $100\times100\,\mu m^2$ in the metallization. In figure\,\ref{LGAD_pics}, two pictures of sensors studied in this work are shown. Also, the electrical characterization of these detectors is shown in figure\,\ref{LGADs_IVCV}, where their main parameters can be extracted. For a better description of the samples these parameters are listed in table\,\ref{table_1}.

\begin{table}[!t]
\begin{center}
\begin{tabular}{| l | c | c | c | c | c |}
\hline
Sample & $V_{dep}$ (bulk) & $V_{gl}$ & $V_{bd}\,(20^oC)$ & $C_{end}$ & Measured thickness\\
\hline
HPK & 7.2 V & 53 V & 170 V & 3.6 pF & $48\,\mu m$ \\
CNM & 3.4 V & 40 V & 112 V & 4.2 pF & $42\,\mu m$ \\
\hline
\end{tabular}
\caption{Main parameters of the LGADs used for this work.}
\label{table_1}
\end{center}
\end{table}

\begin{figure}[!t]
\centering
\includegraphics[width=0.45\columnwidth]{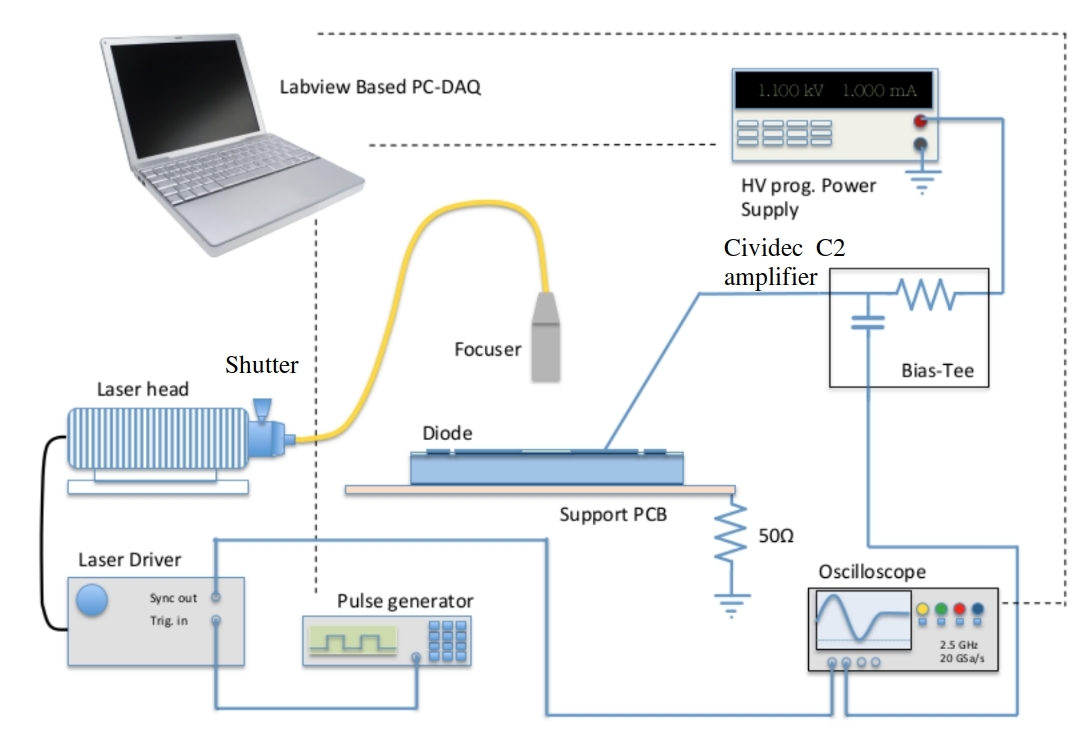}\hfil
\includegraphics[width=0.45\columnwidth]{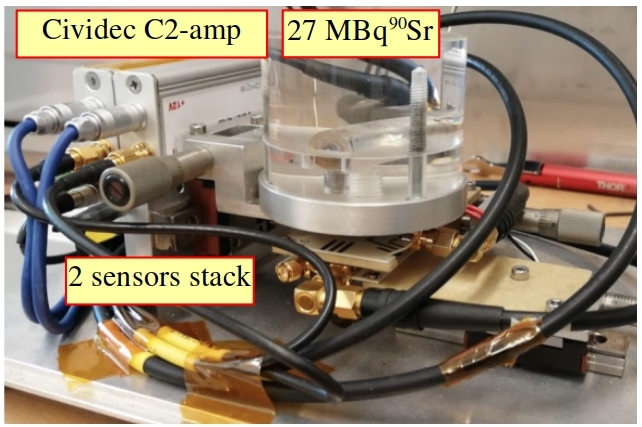}\hfil
\caption{On the left-hand side figure, a schematic view of the TCT set-up with its main components is given. On the right-hand side of the figure, a photo of the radioactive source set-up is shown. The amplifiers, the Sr-90 source and two sensors are indicated.} 
\label{set-ups}
\centering
\end{figure}

The experimental set-ups used for the characterization of the detectors are shown in figure\,\ref{set-ups}.
The transient current technique (TCT), depicted on the left-hand side of the figure, is an important tool to study several detector features like signal formation, charge collection, and trapping mechanisms \cite{Kramberger:1390490}. The basic principle of this technique is to use a laser pulse to generate excess carriers inside the detector bulk, then these carriers drift towards the respective electrodes under the influence of an applied bias voltage while the induced current signal is measured. For the characterization of the devices, a sub-ns pulsed infrared laser (1060 nm) that creates charge carriers all along the optical path was used. All presented data were obtained by  illuminating the devices from the pad side (top illumination).

\bigskip

The main components of the TCT set-up are shown in figure\,\ref{set-ups} on the left-hand side plot. The set-up is placed inside a Faraday cage, the Device Under Test (DUT) is glued on a customized Printed Circuit Board (PCB) that is placed on top of a temperature controlled metallic support. The humidity is controlled by supplying dry air inside the Faraday cage. The laser induced signal is amplified using a Cividec C2 current amplifier (2 GHz, 40 dB). After the amplification stage, the signal is digitized with an Agilent DSO 9254 Oscilloscope (2.5 GHz, 20 Gsa/s). The intensity of the laser can be tuned by using a mechanical attenuator. To correct possible fluctuations of the laser intensity during the measurements, a calibrated reference sensor receives a fraction of the laser beam after a beamsplitter.

\begin{figure}[!t]
\centering
\includegraphics[width=0.45\columnwidth]{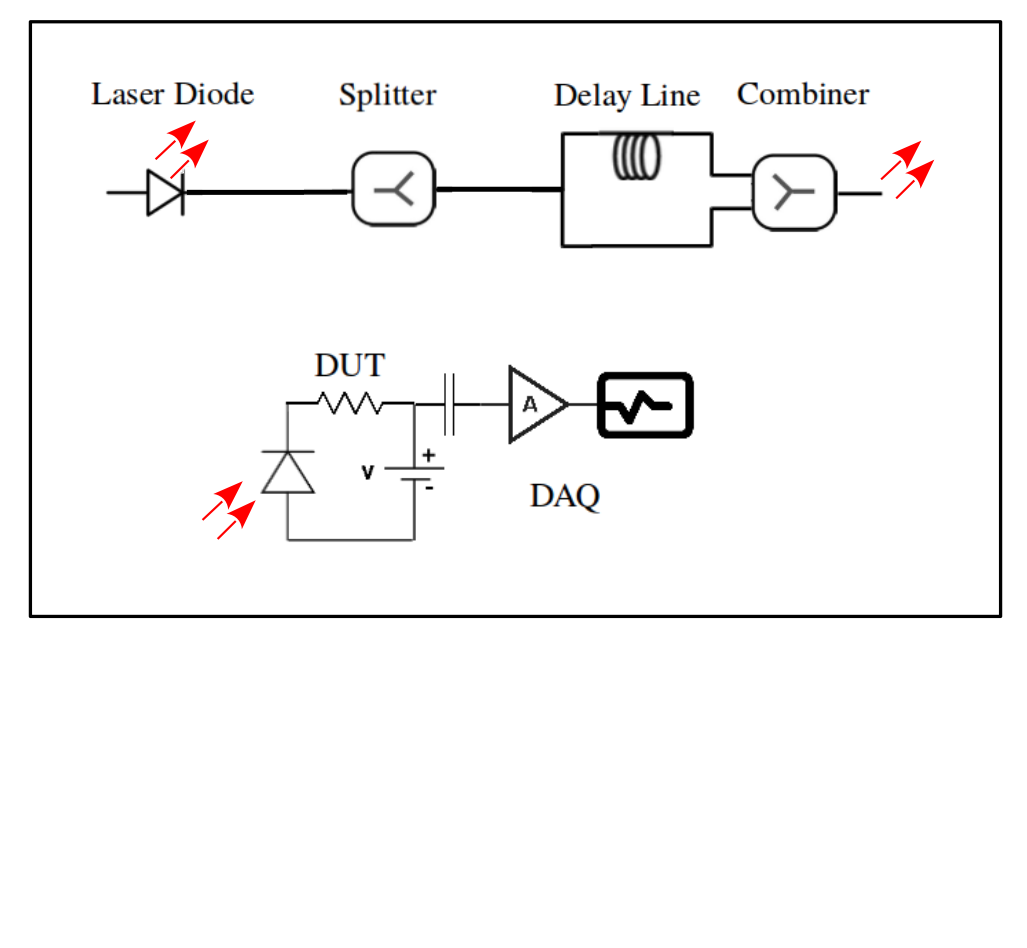}\hfil
\includegraphics[width=0.55\columnwidth]{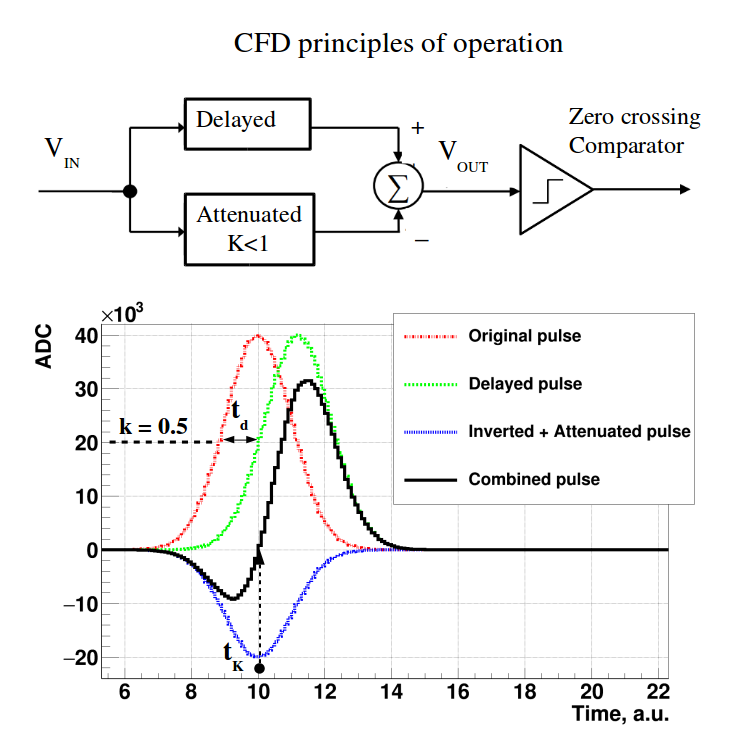}\hfil
\caption{On the left-hand side figure, timing setup schematic with all its main components: laser diode, isolator, splitter, delay line, combiner and DUT. On the right-had side figure, scheme of CFD principle of operation. In this example an attenuation value of k = 0.5 has been chosen. $t_{d}$ represents the time that the original signal is delayed and $t_{k}$ would be the reference time extracted with this method.} 
\label{timing-setup}
\centering
\end{figure}

Timing measurements are possible with the IR-laser too. To perform them, the IR optical line is modified to guide two consecutive laser pulses onto the DUT. In this case, an external reference is not needed and it is possible to measure the intrinsic time resolution of the DUT. The schematic of the modified optical line is shown in figure\,\ref{timing-setup} on the left-hand side plot. Each IR-laser pulse is split into two optical lines, in one of which a fixed delay of $\sim$50\,ns was introduced. Then, these two lines are recombined in one line that illuminates the  DUT. In this way, a fixed time interval between laser pulses arriving at the DUT is introduced. The difference in intensity between the two pulses is about 1.5\%. The IR-laser intensity can be tuned with the shutter to a value corresponding to a generated charge of $\sim1\,MIP$ (Minimum Ionizing Particle). More details about the set-up can be found in \cite{WIEHE2021}.

\bigskip

Using a Sr-90 source it is possible to perform gain and timing measurements with MIPs. The beta source set-up is placed inside a climatic chamber for temperature and humidity control. Also, the detectors are shielded from electromagnetic noise by a customized Faraday Cage inside the climatic chamber. The main components of the set-up can be seen in figure\,\ref{set-ups} on the right-hand side picture. The Sr-90 source is placed on top of a two (or three, when two DUTs are measured at the same time) sensors stack aligned underneath the source. The DUT is always placed on top, closest to the source, and the reference detector on the bottom. The signal generated by the beta particles is amplified using a Cividec C2 current amplifier and further digitized by an Agilent DSO 9254 Oscilloscope. An identical read-out chain is used in the laser and the source set-up. All the sensors present in the stack are included in the trigger algorithm, therefore a signal in all of them is required for each acquired event.

In both set-ups, the time resolution is computed using a Constant Fraction Discrimination algorithm (CFD).

The CFD method used in this paper is a software emulator of a real CFD electronic circuit. CFD is a technique developed to provide information about the arrival time of an event with no dependency on the amplitude of the signal. The principle of operation is based on detecting the zero crossing of a bipolar signal obtained by subtracting a fraction of the input signal ($0 < k < 1$) to its delayed copy as it is illustrated in figure\,\ref{timing-setup} on the right-had side plot. This bipolar signal crosses the baseline at a fixed time ($t_{k}$) with respect to the start of the original signal.

\section{Differences between IR-laser and Sr-90 measurements: the gain damping mechanism}

\begin{figure}[!t]
\centering
\includegraphics[width=0.45\columnwidth]{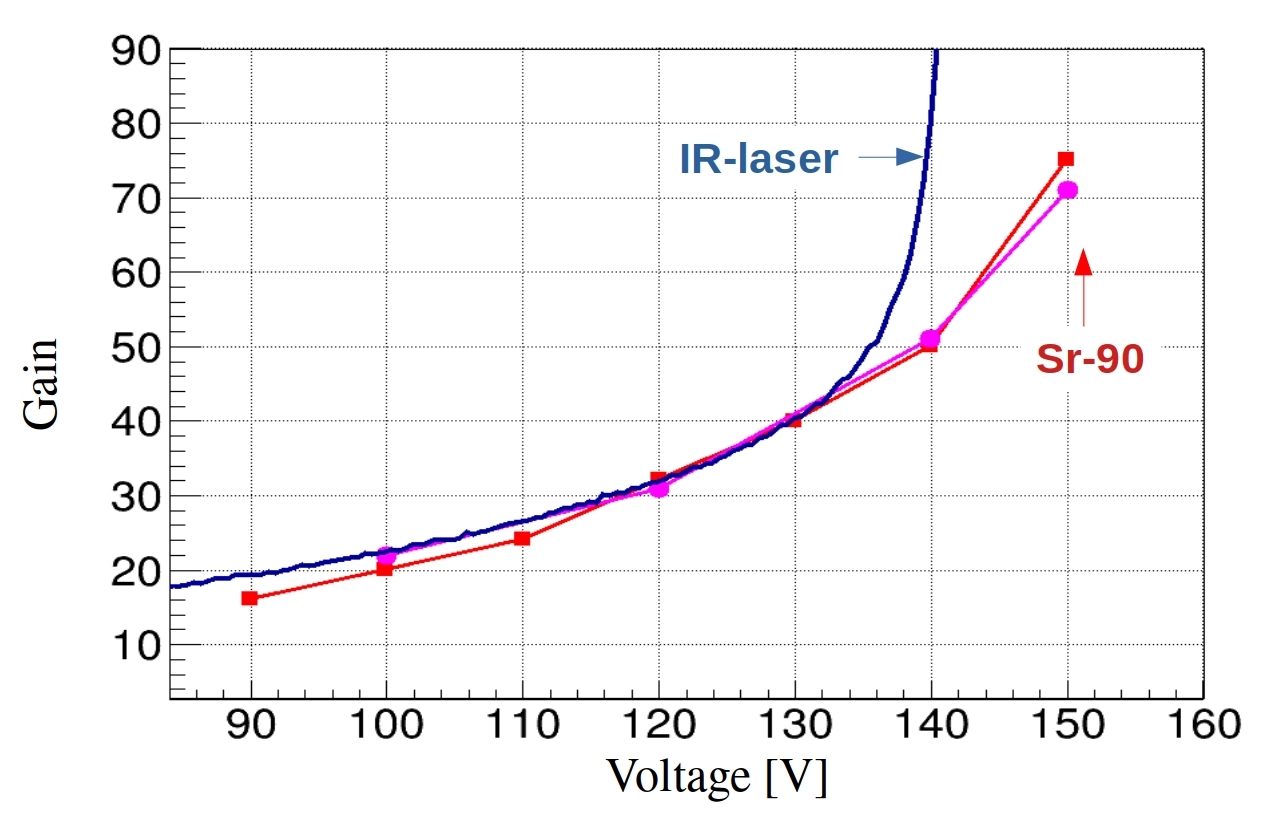}\hfil
\includegraphics[width=0.45\columnwidth]{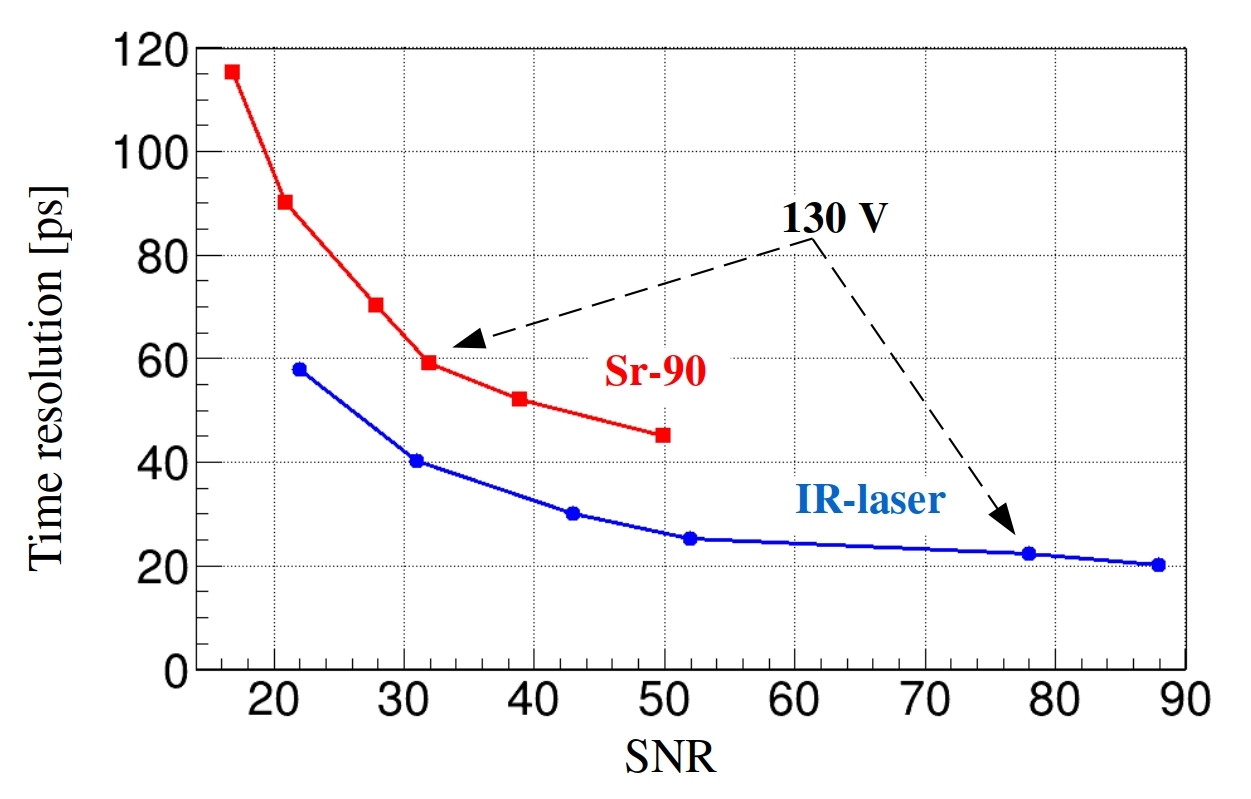}\hfil
\caption{On the left-hand side plot: measured gain as a function of the voltage, with IR-laser and Sr-90 source, for a $50\,\mu m$ LGAD sensor. On the right-hand side plot: for the same LGAD, time resolution as a function of the SNR measured with IR-laser and Sr-90 source. For these measurements, the IR-laser was tuned to $\sim1 MIP$.} 
\label{RS-TCT}
\centering
\end{figure}

The main aim of the presented work is to evaluate how comparable measurements on LGADs taken with an IR-laser and a Sr-90 source are, and to understand the systematic differences when comparing the results obtained from these two techniques. 
For this purpose, in a first step the IR-laser intensity was tuned to provide the same equivalent charge of $\sim1\,MIP$ as the most probable value of charged obtained with the Sr-90 source. This was calibrated with a $50\,\mu m$ PIN detector where the the IR-laser intensity was adjusted to get a total collected charge of 0.54\,fC after full depletion. As the IR-laser absorption is temperature dependant, the calibration was done at $20^oC$. This is the temperature used for all the measurements presented in this paper. This calibration, in principle, should result in comparable sensor characteristics from the two measurements. However, important differences were observed, as shown in figure\,\ref{RS-TCT}, where the measured gain and the time resolution from IR-laser and Sr-90 source measurements are compared for the same detector.The gain is obtained as the ratio between the charge measured with the LGAD and the charge measured with the corresponding PIN, after full depletion, under the same conditions (temperature and laser intensity). On the left-hand side plot of figure\,\ref{RS-TCT}, it is demonstrated that the measured gain, as function of voltage, for the same LGAD shows significant differences between the data obtained with the laser and the data obtained with the Sr-90 source. At  low voltages, up to about 130 V, the IR-laser and source data agree, however, at higher voltages the gain measured with the IR-laser is much higher than the one measured with the SR-90 source.

\bigskip

The time resolution, that is computed in both cases using the previously described CFD method, is shown as a function of the Signal-to-Noise Ratio (SNR) on the right-hand side plot of figure\,\ref{RS-TCT}. The time resolution measured with the Sr-90 source is affected by Landau fluctuations of the deposited charge. This phenomenon does not occur for photons. The measured time resolution obtained with the laser can be interpreted as mainly the intrinsic jitter of the device. For this reason, it is expected to obtain a better time resolution with the IR-laser than with the Sr-90 source for the same SNR. Although for these samples the additional contribution given by the Landau fluctuations was not measured, the time resolution measured with the source is around $\sim40-50\,\%$ worse at low SNR and increases up to $\sim80\,\%$ at higher SNR. This difference is much higher than expected for a $50\,\mu m$ thick detector, where the jitter and Landau noise contributions to the time resolution should be comparable \cite{Sadrozinski_2017}. For example, the measured time resolution from the jitter contribution at 130V is $\sim\,20\,ps$. This implies that the Landau contribution to the time resolution should add $\sim\,20\,ps$ extra, while the measured one with the Sr-90 source (jitter plus Landau noise) at 130V is $\sim\,60\,ps$, $\sim\,50\%$ higher than the expected $\sim\,40\,ps$.

\bigskip

These differences observed between data obtained from IR-laser and Sr-90 source measurements can be interpreted considering that, despite the same amount of charge carriers are generated in the bulk of the detector, in the case of the IR-laser the charge density is lower as the charge is generated inside a larger volume (Gaussian beam waist $\sim\,6\,\mu m$). In the case of the Sr-90 the charge is generated in a much narrower ionizing path, and projecting a higher charge density into the gain layer. Too many charge carriers inside a small gain layer volume will produce a local reduction in the electric field, a screening effect \cite{McKeon:568762}, and this will lead to a reduction in the impact ionization parameter \cite{Maes1990ImpactII}, resulting in a lower measured gain. To prove this hypothesis, IR-laser and Sr-90 source measurements varying the charge density at different bias voltages (electric field in the gain layer) were performed.

\section{IR-laser measurements}

To proof the previously described gain damping mechanism in LGADs, the charge density generated inside the bulk was varied. For this end, two different types of measurements were performed.

\subsection{Variation of the laser intensity with constant focus}

\begin{figure}[t]
\centering
\includegraphics[width=0.326\columnwidth]{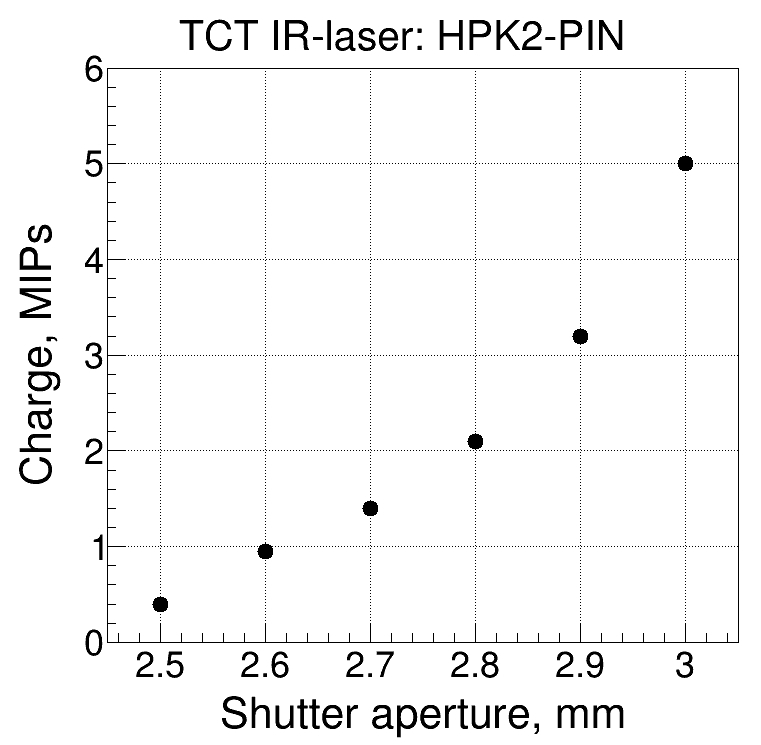}\hfil
\includegraphics[width=0.332\columnwidth]{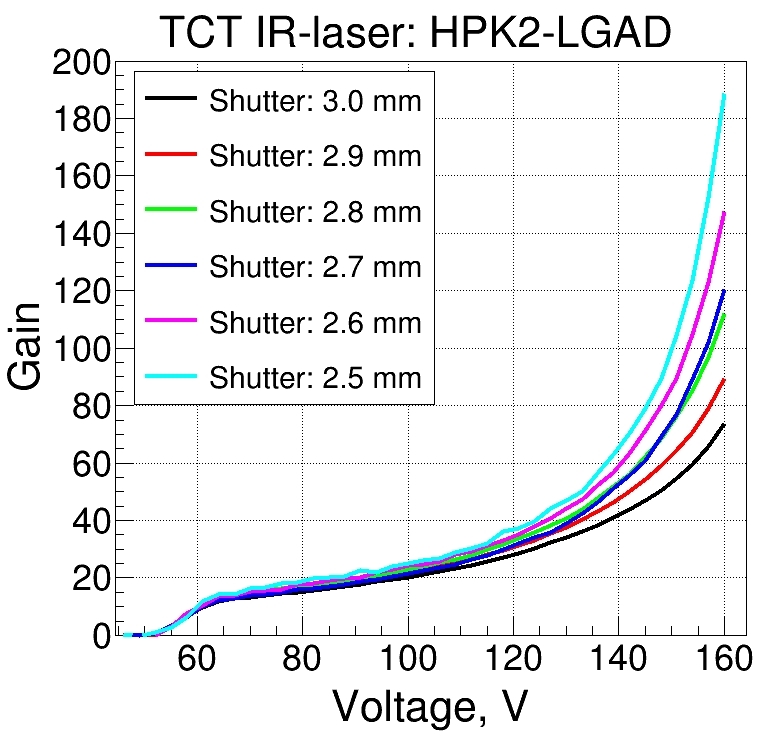}\hfil
\includegraphics[width=0.33\columnwidth]{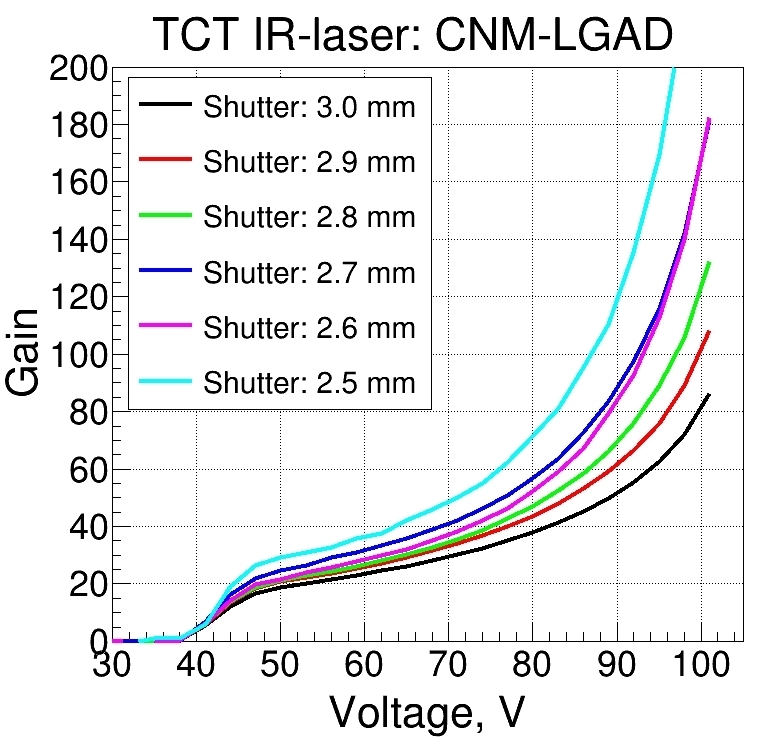}\hfil
\caption{On the left-hand side figure, charge measured in an HPK PIN detector for different shutter apertures. The charge measured is shown in MIPs. On the middle and right-hand side figures: gain measured as a function of the bias voltage for different IR-laser intensities for the HPK and CNM LGADs respectively.} 
\label{IR-laser-shutter}
\centering
\end{figure}

It is possible to modify the laser intensity by opening or closing the laser shutter while the laser spot size in the sensor remains the same, $\sim10\,\mu m$ in FWHM when focus. Like this, the charge density inside the detector bulk, that is the driver of the damping mechanism, is being modified. For 1 MIP, and assuming a cylindrical ionizing volume of $\sim10\,\mu m$ in diameter and $\sim50\,\mu m$ in length, the charge density inside the bulk will be $\sim1\,e^-\mu m^{-3}$: The measurements at different laser intensities are divided into two sets: for the high-intensity ones the amplifier was not used to avoid saturation on the signal, and for the low-intensity measurements the amplifier was used. In this last case, the pulse height after amplification of the TCT signal was always below 1 V (limit of the linear range of the amplifier). The results for the low-intensity measurements are shown in figure\,\ref{IR-laser-shutter}. This was measured for values between $\sim$0.5 MIPs to $\sim$30 MIPs. The laser intensity was calibrated for a PIN diode. Figure\,\ref{IR-laser-shutter} (left) shows  the conversion between the shutter opening and the equivalent number of MIPs induced in the detector. An averaging of signal was done to improve the SNR in the gain measurements.

\subsection{Variation of the focus with constant laser intensity}

\begin{figure}[!t]
\centering
\includegraphics[width=0.99\columnwidth]{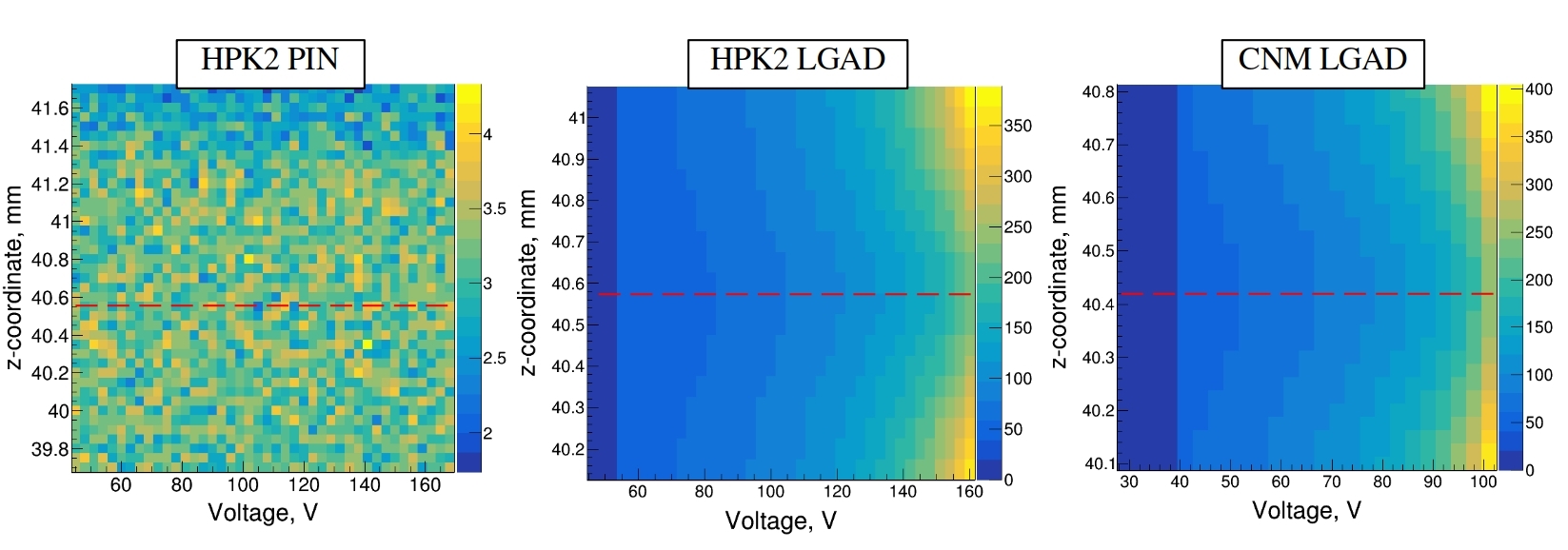}\hfil
\caption{Out-of-focus scans as a function of the reverse bias for one HPK PIN detector (left), one HPK LGAD (center), and one CNM LGAD (right). The focus position in the z-coordinate is indicated by a red dotted line. The laser spot was always inside the opening window on the metallization in all cases.} 
\label{IR-laser-focus}
\centering
\end{figure}

The second type of measurements were done by moving the sample out of the focus position while keeping the laser intensity constant. When the IR-laser is focused inside the detector, it is generating a higher charge density than when it is not focused. The total charge generated inside the sensor, however, remains the same, only the volume illuminated for the same amount of charge is different. These out-of-focus measurements were performed by moving the sample with respect to the focal point (z-coordinate in the set-up). The results of these out-of-focus measurements can be seen in figure\,\ref{IR-laser-focus}. The focal position of the IR-laser is indicated by the red dotted line in the plots. In the colormap palette the measured charge in fC is shown. The measurements were performed at $20^oC$ and with an IR-laser shutter aperture of 3.1 mm, that is equivalent to $\sim7\,MIPs$. The PIN device is fully depleted at around 10 V. 

\section{Characterization with radioactive source}

One way to study the effect of the charge density in the measured gain with the Sr-90 source is by tilting the DUT from normal incidence (at $0^o$). When the DUT is tilted a small angle ($\alpha$), as a first-order approximation, the most probable value of the collected charge will increase by a factor $\beta$, see equation \ref{eqn_1}. This is due to the larger ionizing path of the electrons inside the bulk. There is also an impact on the statistical distribution of the deposited charge \cite{Bichsel:RevModPhys.60.663}, but this is assumed to be negligible when changing the effective sensor thickness from 50 $\mu$m to a value of 51.5 $\mu$m corresponding to the effective length of the maximum tilting angle.
To perform these measurements, the DUT was tilted by: $7^o$ and $14^o$. In the case of $7^o$ the increase in charge with respect to the $0^o$ position is $\beta\,=\,\sim\,1.01$ and for $14^o$ it increases in a factor $\beta\,=\,\sim\,1.03$.

\bigskip
\begin{equation} \label{eqn_1}
\beta = \frac{1}{cos(\alpha)}
\end{equation}

\bigskip
Tilting the DUT does not modify the charge density along the ionizing path in the bulk in a significant way, but it modifies the charge density of the charges arriving to the gain layer that is what triggers the gain damping mechanism. In other words, the projected charge density under the gain layer decreases proportionally a factor $\gamma$, given by equation \ref{eqn_2}, where 'd' is the active thickness of the detector. 

\bigskip
\begin{equation} \label{eqn_2}
\gamma = \frac{d}{sin(\alpha)}
\end{equation}

\begin{figure}[t]
\centering
\includegraphics[width=0.95\columnwidth]{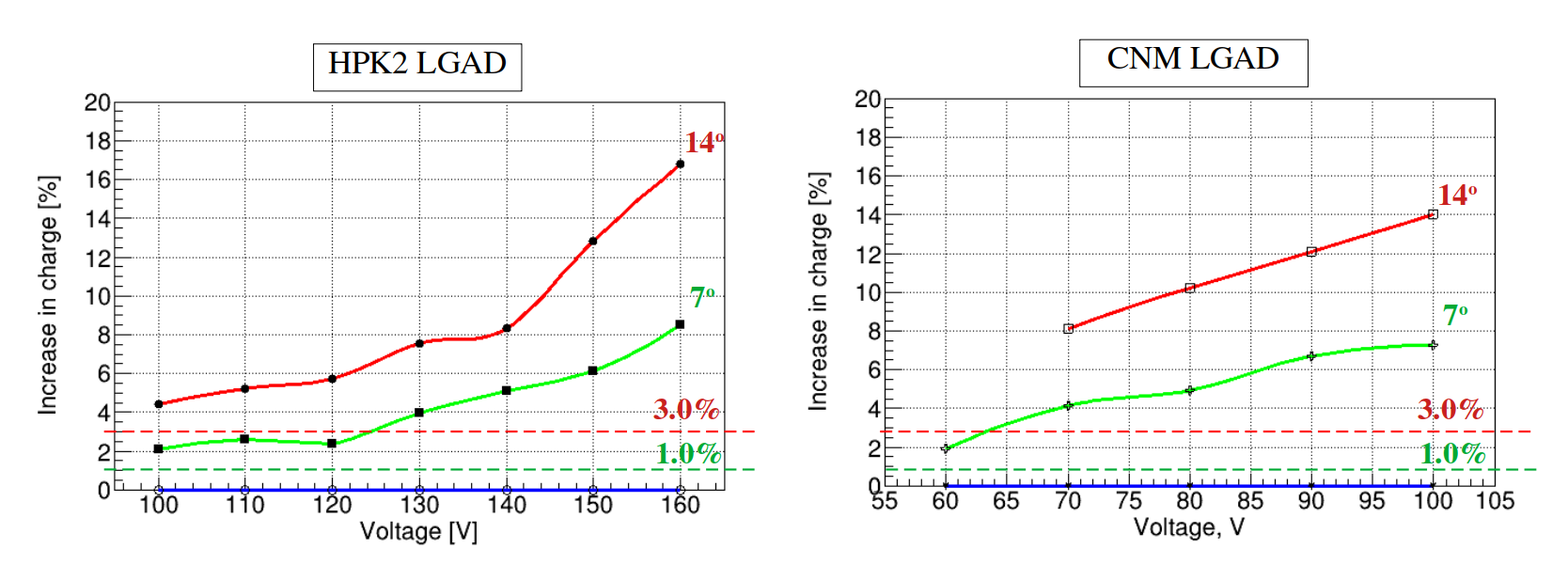}\hfil
\caption{Increase in the most probable value of the measured charge, with respect to the $0^o$ position, as a function of the bias voltage for different rotation angles. The horizontal dotted lines show the expected increase in the collected charge due to the increased ionization path length by rotating the DUT.} 
\label{SR90-charge-angle}
\centering
\end{figure}

\bigskip
Measurements were performed at $20^oC$ for two different DUTs. The reference detector was an HPK LGAD from the same production run as the HPK sensor under test, but with lower gain. The bias voltage was kept constant for the reference detector throughout the  whole experiment (180 V). The trigger was set in coincidence between the DUT and the reference detector, and 20.000 events were collected for each bias voltage. The charge distribution was fitted using a Landau function convoluted with a Gaussian function. The charge value considered for the gain measurements was the Most Probable Value (MPV) of the fit.  Figure \ref{SR90-charge-angle} shows the increase in the measured charge with respect to the $0^o$ position for one HPK LGAD and one CNM LGAD.
The PIN detector was measured using a charge sensitive amplifier (CSA), obtaining a MPV of $\sim\,0.54\,fC$. This results is in agreement with the expectations for a $48\,\mu m$ thick sensor \cite{CurrasRivera:2291517}. The CSA amplifiers present better SNR than current amplifiers, but due to their large shaping time, they are not suitable for timing applications. 

\bigskip

\begin{figure}[t]
\centering
\includegraphics[width=0.95\columnwidth]{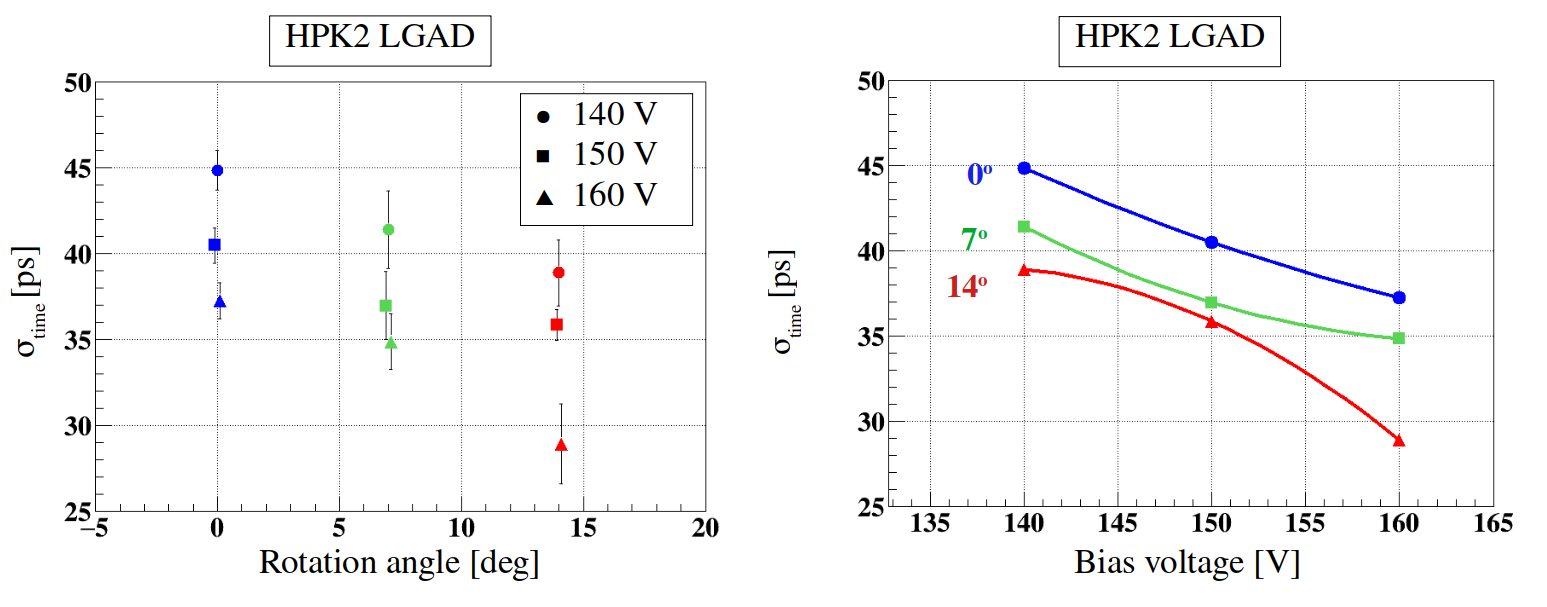}\hfil
\caption{Time resolution measured for the three angle configurations in one HPK LGAD. On the left is shown as a function of the rotation angle and on the right as a function of the reverse bias.} 
\label{SR90-time-angle}
\centering
\end{figure}

Once the effect in the collected charge was measured, the set-up was set into a three samples configuration to be able to extract the time resolution with more precision and see if the tilting of the DUT translates in an improvement in the time resolution too. Following the same approach as before, but this time with one extra reference sensor (both identical and biased at 180 V), the time resolution was investigated  tilting the DUT to the same angles as before: $0^o$, $7^o$ and $14^o$. The time resolution measured for the three angle configurations is shown in figure \ref{SR90-time-angle}, on the left-hand side plot as a function of the rotation angle and, on the right-hand side plot as a function of the reverse bias. 

\bigskip

\clearpage % push out all figures before next section
\section{Discussion}

In figure\,\ref{RS-TCT} differences between IR-laser and Sr-90 source measurements in LGADs were presented  summarizing the main problem that this work is addressing. The observed differences in gain and in time resolution can be explained by the different charge densities generated by the two different ionizing processes induced by the IR-laser or the Sr-90 source. Considering only normal incidence for simplicity in the explanation, this process can be understood as follows:  the higher the carrier density generated in the bulk, the greater the carrier density arriving at the gain layer. As soon as the carriers move into the gain layer, the high electric field leads to impact ionization and the number of carriers quickly gets multiplied. At this point, the free charge carrier density inside the gain layer keeps increasing, largely surpassing  the one initially generated in the bulk. This high concentration of charge carriers inside a very small volume in the gain layer produces a local screening effect, i.e. a local reduction of the electric field strength. The reduced field strength in turn, decreases the efficiency of the impact ionization process, resulting in a reduction of the total charge generated by impact ionization in the gain layer. The latter being the value that determines the measured gain of the devices. It is important to remark that the impact ionization process has an exponential dependence on the electric field strength. Therefore, the absolute value of the gain reduction  is strongly affected by both, the strength of the electric field in the gain layer and the amount of charge generated by impact ionization. This behavior is reported in figure\,\ref{RS-TCT} (left), where the gain curve from IR-laser measurements increases with the reverse bias more rapidly than the gain curve from Sr-90 source measurements; that is the process with higher ionizing charge density generation. 

\bigskip

This described "gain suppression" mechanism was proven with IR-laser by modifying the charge density in two different ways that lead to the same conclusion: The higher the charge density of the charge carriers entering the gain layer, the lower the measured gain in the LGAD. It can be seen in figure\,\ref{IR-laser-shutter}, that by increasing the laser intensity the gain measured in the LGAD decreases. This is observed for both types of LGADs used in this work: the devices produced by HPK with a deep implanted gain layer and the devices from CNM with a gain layer attached to the $n^+$-implant. The effect is especially relevant at high voltages, where the electric field in the gain layer is higher and the damping mechanism is more pronounced due to the exponential dependence of the impact ionization coefficients on the electric field strength. In the same way, the focus variation measurements show similar behavior, as reported in figure\,\ref{IR-laser-focus}. The highest charge density is achieved when the laser is focused and the charge density is decreasing when the detector is moved in the z-coordinate away from the focal position.  In the case of the PIN detector, the charge measured as a function of the z-coordinate is not changing, and is independent of the applied reverse bias voltage. In the case of the LGADs, the amount of collected charge highly depends on the z-coordinate and the applied voltage. It is also observed that the impact of the de-focussing is enhanced at high voltages where the electric field strength in the gain layer is higher.

\bigskip

Measurements with the Sr-90 source tilting the DUTs, and in consequence modifying the charge density entering the gain layer, lead to the same conclusions. Figure\,\ref{SR90-charge-angle} shows a clear increase in the measured charge with respect to the standard DUT configuration at 0 degrees. For the two LGADs studied here, the increase in charge is more dominant at high voltages, which is in agreement with the IR-laser measurements. Also, the increase in charge is always above the $\beta$ factor calculated, see equation \ref{eqn_1}, for the two measured angles and plotted in dotted lines in the figures as a reference. Also, in figure \ref{SR90-time-angle} a clear improvement of the time resolution is observed when the position of the DUT changes from $0^o$ to $7^o$, and the tendency continues for the $14^o$ position. This opens a possibility to improve the operational timing performance of these devices. The best timing performance will be reached for non orthogonal tracks, where the charge density under the gain layer is minimized. One possible way to achieve this would be the use of the 3D-detectors structure design, where the electrodes penetrate the sensor bulk perpendicular to the surface. Excellent timing results using standard 3D-detector were already reported \cite{KRAMBERGER:201926,Ugobono:2747755}. If the addition of an internal gain layer along the electrodes should proof feasible, the timing performance could be further enhanced and the sensor should be less impacted by the gain suppression mechanism presented in this work.

\bigskip

To conclude, for the testing and qualification of detectors with internal gain, the conditions in which the gain is evaluated must be carefully chosen. It is not straightforward to compare results obtained in experiments with different ionization densities, e.g. comparing laser tests against particle beam experiments. Therefore, the testing conditions should be chosen to be as close as possible to the one in the foreseen application of the sensors. In case of laser experiments, the ionization density has to be reported to allow for a proper comparison between different experiments. One of the main points that it is stressed in this work is that the reported gain in LGADs, and linked to this the reported charge and timing capabilities, are significantly affected by the experimental method used to generate the data.  

\section{Summary}

It was demonstrated in this work that the total amount of collected charge in LGAD sensors, and thus the measured gain and timing performance of the device, depends on the ionization density of the deposited charge in the detector. The observed effect leads to a suppression of the gain for high ionization densities. Experimental data of gain measurements on 50 $\mu$m thick LGAD sensors obtained with infrared laser pulses and Sr-90 beta particles were presented and in both cases the gain suppression effect was observed. An important outcome is that it is not straightforward to compare results obtained by laser and source measurements, as the ionization density in the two cases differ significantly. This is especially relevant for the evaluation of the time resolution of LGAD detectors. Not only does the statistical distribution of the deposited charges differ between source and laser measurements (i.e. with/without Landau fluctuations), but also does the LGAD gain differ even if the same amount of charge has been deposited. 
The ionization density is therefore an important parameter to be taken into account in evaluating and operating LGAD detectors. 

\section*{Acknowledgement}

This work was performed in the framework of the RD50 Collaboration and the CERN $EP-R\&D$ Programme on Technologies for Future Experiments.

\bigskip
\pagebreak[1]

\bibliography{mybibfile}

\end{document}